\documentclass[12pt]{article}

\usepackage{amsmath}
\usepackage{amssymb}
\usepackage{graphicx}

\title{Waveform Selectivity at the Same Frequency}

\author
{Hiroki Wakatsuchi,$^{1,2\ast}$ Daisuke Anzai,$^2$ Jeremiah J. Rushton,$^3$ \\Fei Gao,$^{4,3}$ Sanghoon Kim,$^3$ Daniel F. Sievenpiper$^3$\\
\\
\normalsize{$^{1}$Center for Innovative Young Researchers,}\\
\normalsize{$^{2}$Department of Electrical and Electronic Engineering,}\\
\normalsize{Nagoya Institute of Technology, Gokiso-cho, Showa, Nagoya, Aichi, 466-8555, Japan}\\
\normalsize{$^{3}$Applied Electromagnetics Group, Electrical and Computer Engineering Department,}\\
\normalsize{University of California, San Diego, 9500 Gilman Drive, La Jolla, CA 92093, USA}\\
\normalsize{$^{4}$The Science and Technology on Antenna and Microwave Laboratory,}\\
\normalsize{Xidian University, Xi$\acute{~}$an, Shaanxi 710071, China}\\
\\
\normalsize{$^\ast$E-mail: hirokiwaka@gmail.com}
}

\date{}

\begin{document} 




\maketitle


\hyphenpenalty=1000

\textbf{Electromagnetic properties depend on the composition of materials, i.e.\ either angstrom scales of molecules or, for metamaterials, subwavelength periodic structures. Each material behaves differently in accordance with the frequency of an incoming electromagnetic wave due to the frequency dispersion or the resonance of the periodic structures. This indicates that if the frequency is fixed, the material always responds in the same manner unless it has nonlinearity. However, such nonlinearity is controlled by the magnitude of the incoming wave or other bias. Therefore, it is difficult to distinguish different incoming waves at the same frequency. Here we present a new concept of circuit-based metasurfaces to selectively absorb or transmit specific types of waveforms even at the same frequency. The metasurfaces, integrated with schottky diodes as well as either capacitors or inductors, selectively absorb short or long pulses, respectively. The two types of the circuit elements are then combined to absorb or transmit specific waveforms in between. This waveform selectivity gives us another freedom to control electromagnetic waves in various fields including wireless communications, as our simulation reveals that the metasurfaces are capable of varying bit error rates in response to waveforms.
}
\clearpage


\begin{figure*}[t!]
\centering
\includegraphics[width=0.95\textwidth]{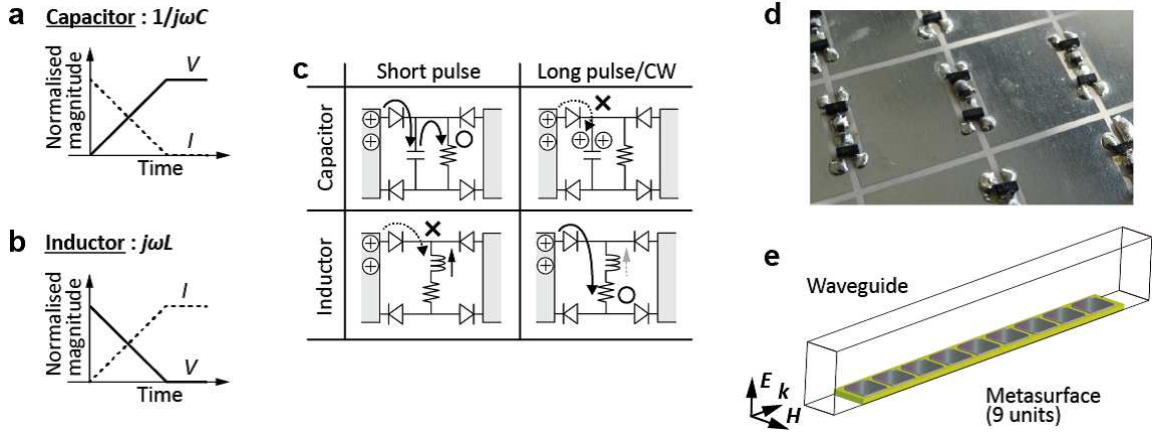}
\caption{\label{fig:models} \textbf{Fundamental concept of waveform-dependent metasurfaces. a} and \textbf{b,} Time domain responses of \textbf{(a)} a capacitor and \textbf{(b)} inductor to a rectified signal. \textbf{c,} Use of a resistor between metasurface patches leads to absorption of either a short or long pulse. \textbf{d,} One of measured samples (inductor-based metasurface). Each circuit element was soldered and electrically connected to the surface. \textbf{e,} Metasurfaces were built on a 1.52 mm height dielectric substrate (Rogers 3003) and simulated in a TEM waveguide (22 mm tall and 18 mm wide). See Supplementary Information for more detail.}
\end{figure*}

The advent of new properties has been leading to the development of new applications in electromagnetism. For example, metamaterials \cite{smithDNG1D,smithDNG2D2,EBGdevelopment}, composed of sub-wavelength resonant structures, readily enable us to use negative- or zero-refractive indices \cite{smithDNG2D2} as well as artificially engineered high impedance surfaces \cite{EBGdevelopment}. These unusual properties were applied to the development of a diffraction-limitless lens \cite{pendryperfetLenses,fangSuperlens}, cloaking device \cite{pendryCloaking,enghetaCloaking,fridman2012demonstration}, unusually thin absorbers ($<$$\lambda$/4 where $\lambda$ is the wavelength of the incoming wave) \cite{mtmAbsPRLpadilla,My1stAbsPaper,aplNonlinearMetasurface}, etc. Importantly, such new electromagnetic properties and applications were widely exploited in other disciplines such as the acoustics \cite{fang2006ultrasonic,torrent2007acoustic,li2009experimental,popa2011experimental}, thermodynamics \cite{han2013homogeneous} and vibration engineering \cite{ding2007metamaterial,farhat2009ultrabroadband}. Moreover, a recent study on metasurface absorbers containing circuit elements such as diodes and capacitors demonstrated a new property called \emph{waveform dependence} \cite{wakatsuchi2013waveform,wakatsuchi2013experimental}. Interestingly these metasurfaces absorbed only short sine wave pulses, while transmitting continuous waves (CWs) even at the same frequency. In this study we present a new concept of metasurfaces to fully control waveforms, i.e.\ selective absorption of either short pulses or CWs as well as absorption or transmission of specific waveforms in between. This new property termed \emph{waveform selectivity} allows us to distinguish incoming waves in an unusual manner depending not only on the frequency but also on the waveform. Thus the waveform selectivity has a potential to develop new kinds of techniques and applications in electromagnetism as well as in other disciplines. 

The waveform selectivity is made possible by integrating the rectification of microwave diodes with time-domain responses of capacitors and inductors (Figs.\ \ref{fig:models}(a) and \ref{fig:models}(b)). First, a diode converts the frequency of an incoming signal to an infinite set of frequency components but mainly to zero frequency \cite{My1stAbsPaper}. The conversion to zero frequency is further enhanced, if a four-diode bridge is used \cite{wakatsuchi2013waveform} (see Supplementary Information in detail). Also, a capacitor has an impedance calculated from $1/j\omega C$, where $j^2=-1$, $\omega$ is the angular frequency ($\omega=2\pi f$ and $f$ is the frequency) and $C$ is the capacitance. Since the rectified signal contains the zero frequency component, the capacitor stores the energy but is gradually charged up. This indicates that capacitors allow current to come in \emph{during} an initial time period. On the other hand, an inductor has impedance of $j\omega L$, where $L$ is the inductance, and generates an electromotive force toward the incoming current. This force, however, is gradually weakened due to the zero frequency component, and more current comes in. Thus, inductors accept incoming current \emph{after} an initial time period. 

In addition, integration of these time domain responses with a resistor leads to effective absorption of either short or long waveforms. Fig.\ \ref{fig:models}(c) illustrates such circuit configurations deployed between square patches of metasurfaces. In this figure a capacitor is connected to a resistor in parallel, while an inductor in series. Under these circumstances the capacitor is capable of fully storing the energy of a short pulse during the illumination, and discharges it into the resistor, which dissipates the energy before a next short pulse comes in. For a CW, however, the capacitor is fully charged up. As a result, the incoming wave transmits over the metasurface. Regarding the inductor, a short pulse cannot be rectified by diodes due to the presence of the electromotive force, resulting in no energy dissipation in the series resistor. However, reduction of the force due to long-term illumination permits the rectification, leading to energy dissipation in the resistor. For these reasons a short pulse or CW is expected to be absorbed by the combination of a resistor with a capacitor or inductor, respectively. 

Importantly, selective absorptions and transmissions demonstrated below are not due to the variation in the bandwidth of the incoming frequency, since the pulse widths used are long enough. In other words, the bandwidth of the signals is small compared to that of the surface without any nonlinear elements (e.g.\ provided that frequency and pulse width are respectively 4.0 GHz and 50 ns as set below, the one cycle is 0.25 ns and corresponds to only a 200th of the pulse duration). 

\begin{figure*}[t!]
\centering
\includegraphics[width=0.95\textwidth]{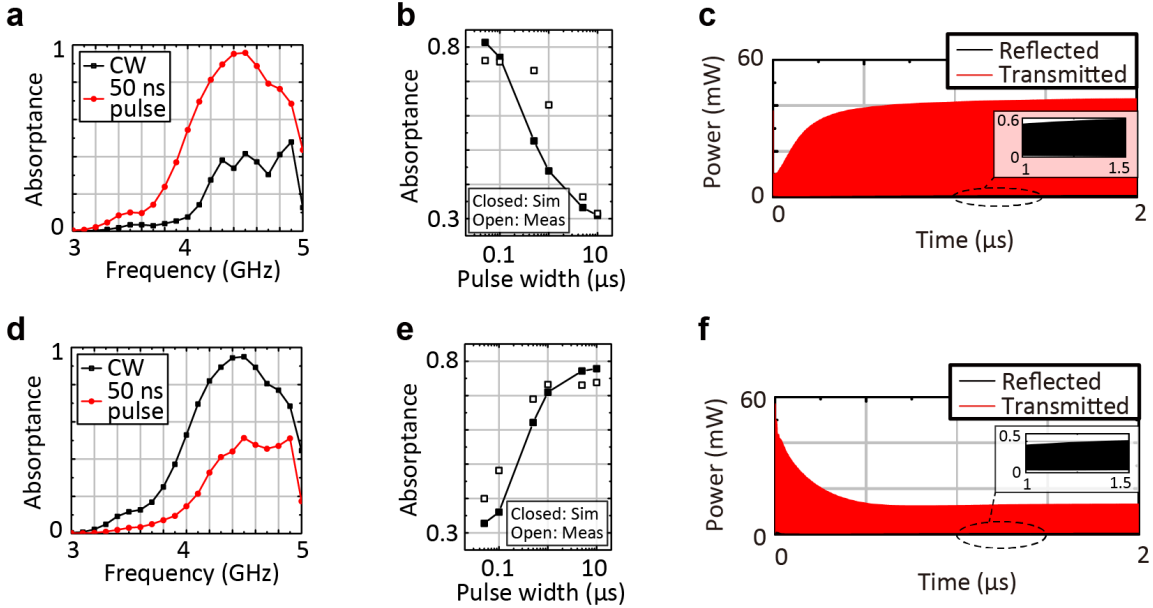}
\caption{\label{fig:either} Absorbing performance of the capacitor-based and inductor-based metasurfaces. (a) and (d) The capacitor- and inductor-based metasurfaces containing the circuit configurations of Fig.\ \ref{fig:models} (c) numerically exhibited stronger absorption for a short pulse (50 ns long) and CW, respectively. (b) and (e) At 4.2 GHz each of the capacitor- and inductor-based metasurfaces numerically showed a clear transition between the short pulse and CW, as plotted by the closed squares of (b) and (e), respectively. Similar trends were experimentally demonstrated at 4.0 GHz (the open squares). (c) and (f) The time domain responses of these metasurfaces are seen in (c) and (f), respectively. }
\end{figure*}

On the basis of this theory, a metasurface was designed and built up as in Figs.\ \ref{fig:models}(d) and \ref{fig:models}(e) in order to absorb incoming surface waves. The simulation model had nine periodic unit cells along the propagation direction of a surface wave on the bottom of a TEM (transverse electromagnetic) waveguide. The measurement sample was fabricated under the same circumstances but with a few differences. For example, since the measurement required to use TE (transverse electric) waveguides (WR284 for up to 3.95 GHz and WR187 from 3.95 GHz) as realistic waveguides, the measurement sample had several unit cells along not only the direction of the wave propagation (\textbf{\textit{k}} in Fig.\ \ref{fig:models}(e)) but also that of the incident magnetic field (\textbf{\textit{H}}) to fully occupy the bottoms of the waveguides. In addition, the measurement used commercial schottky diodes (Avago; High Frequency Detector Diodes HSMS-2863/2864), while the simulation used a SPICE model. The capacitors and inductors respectively had capacitance $C=$ 1 nF and inductance $L=$ 100 $\mu$H, while the resistors used with the capacitors and inductors respectively had resistances $R_c=$ 10 k$\Omega$ and $R_l=$ 5.5 $\Omega$, where the measurement sample of the inductor-based metasurface did not use resistor chips, since the inductor chips already contained some resistance value, which was the same as $R_l$. The self-resonant frequencies of the capacitor and inductor chips, $f_c$ and $f_l$, were respectively 300 and 10 MHz. Note that the time constants of the capacitor-based and inductor-based metasurfaces, determined by $R_cC$ and $L/R_l$, were respectively 10 $\mu$s and $\sim$18 $\mu$s. More details on the simulation and measurement are described in Fig.\ \ref{fig:models} and Supplementary Information. 

First of all, a capacitor-based metasurface and inductor-based metasurface were numerically tested with 15 dBm signals at different frequencies as shown in Figs.\ \ref{fig:either}(a) and \ref{fig:either}(d), respectively. As a result, the capacitor-based and inductor-based metasurfaces respectively absorbed short pulses (50 ns long) and CWs more effectively. Next, the frequency was fixed at 4.2 GHz, where either a short pulse or CW was strongly absorbed, while the other was weakly absorbed. Under this circumstance, the capacitor-based metasurface demonstrated a clear transition between the short pulse and CW as plotted by the closed squares of Fig.\ \ref{fig:either}(b) \cite{wakatsuchi2013waveform}. This is because the capacitors used were gradually charged up, as the pulse width increased. The measurement was performed under the same circumstances except the input frequency set to 4.0 GHz due to a minor shift of the entire feature to a lower frequency region (see Fig. 11 of Supplementary Information). As plotted by the open squares of Fig.\ \ref{fig:either}(b), the measurement result also demonstrated that the absorbing performance decreased, as the pulse width increased. On the other hand, the inductor-based metasurface gradually enhanced the absorbing performance by increasing the pulse width as Fig.\ \ref{fig:either}(e), since the electromotive force was weakened and the incoming rectified energy was dissipated with the series resistors. The reasons of the differences between these simulated and measured results are explained by circuit parasitics (e.g.\ extra capacitances in diodes) and superimposed direct current characteristic. The former reason increased the time constant of the capacitor-based metasurface, while the latter decreased that of the inductor-based metasurface. The time domain responses of these two waveform-dependent metasurfaces to a 15 dBm CWs are seen in Figs.\ \ref{fig:either}(c) and \ref{fig:either}(f), respectively. As expected from Figs.\ \ref{fig:either}(b) and \ref{fig:either}(e), the capacitor-based (inductor-based) metasurface gradually increased (decreased) the transmitted power. These plots also show that the reflected powers were limited. 

\begin{figure*}[t!]
\centering
\includegraphics[width=0.95\textwidth]{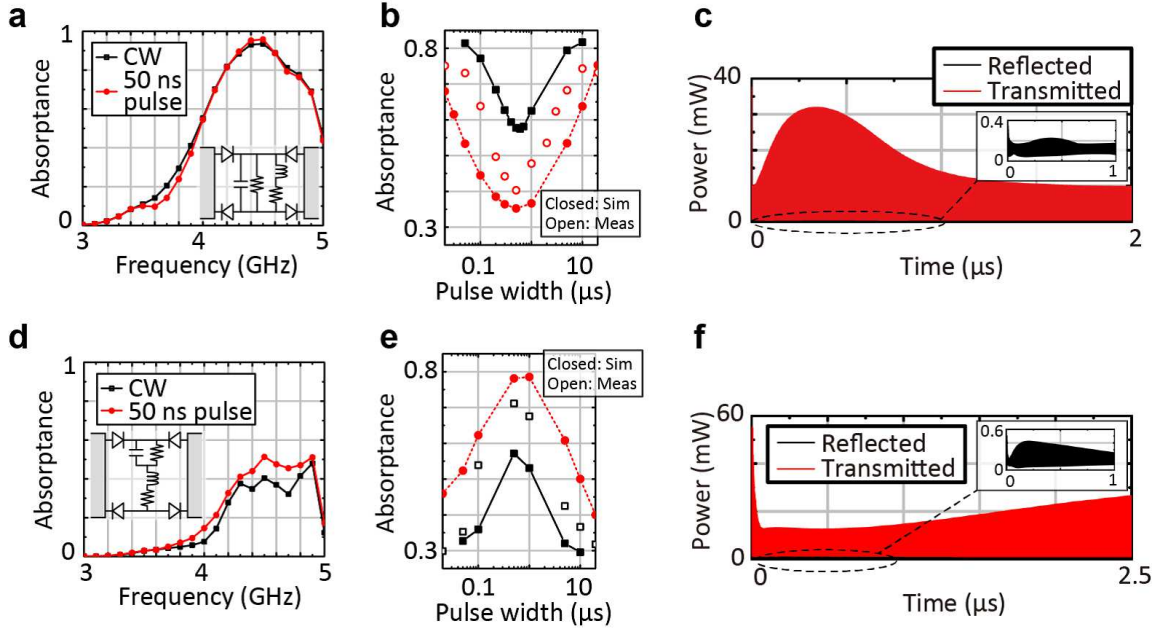}
\caption{\label{fig:both} Selective transmission and absorption achieved through use of both capacitors and inductors. (a) and (d) The insets illustrate the metasurfaces combining both circuit configurations, respectively, (a) in parallel and (d) in series. The parallel-type metasurface numerically shows strong absorption for both of a short pulse and CW, while the series-type shows limited absorption for both of them. (b) The parallel-type metasurface selectively transmits the waveforms which were weakly absorbed by both of the individual capacitor-based and inductor-based metasurfaces in Fig.\ \ref{fig:either} (see the closed squares in Fig.\ \ref{fig:both}(b)). The variation of the absorptance can be controlled by modifying the time constants of the individual metasurfaces, i.e.\ $R_cC$ and $L/R_l$ (the closed circles). The measurement result supports the feasibility of the numerical simulation as well (the open circles). (c) The time domain response of the parallel-type metasurface (the closed squares of Fig.\ \ref{fig:both}(b)) is plotted. (e) In contrast, the series-type metasurface absorbs the waveforms which were strongly absorbed by both of the individual metasurfaces in Fig.\ \ref{fig:either} (see the closed squares in Fig.\ \ref{fig:both}(e)). Despite difference due to circuit parasitics and superimposed direct current characteristics, such a waveform selectivity is experimentally realisable (the open squares). Similarly with the parallel type, the variation of the absorptance can be controlled by the time constants (the circles). (f) The time domain response of the series-type metasurface (the circles of Fig.\ \ref{fig:both}(e)) is plotted. }
\end{figure*}

Such a waveform dependence can be more flexibly designed by combining each of the circuit configurations with another. The insets of Figs.\ \ref{fig:both}(a) and \ref{fig:both}(d) illustrate circuit configurations containing the two types of circuit elements either in parallel or in series. For example, previously the individual capacitor-based metasurface absorbed short pulses but transmitted long pulses, which can be now absorbed by the parallel inductor part as demonstrated in Fig.\ \ref{fig:both}(a). Interestingly, however, such a metasurface can still transmit some waveforms that were weakly absorbed by both of the individual metasurfaces (the closed squares of Fig.\ \ref{fig:both}(b)). In this case, the absorptance value at each pulse width is close to the \emph{larger} value of the individual structures (cf.\ Figs.\ \ref{fig:either}(b) and \ref{fig:either}(e)). Besides, as plotted by the closed circles of Fig.\ \ref{fig:both}(b), the variation of the absorptance curve can be controlled by varying the time constants $R_cC$ and $L/R_l$ (now $C=$ 100 pF, $L=$ 1 mH, $R_c=$ 10 k$\Omega$, $R_l=$ 31.2 $\Omega$, $f_c=$ 1.02 GHz and $f_l=$ 2.4 MHz), since these parameters determine the saturation of each curve. This was experimentally demonstrated by the open circles of Fig.\ \ref{fig:both}(b), where an incoming wave was more transmitted, when the pulse width was around 0.5 $\mu$s. The differences between the measured and simulated results were mainly due to some circuit parasitics and the superimposed direct current characteristics as explained above. Such a behaviour of the parallel-type metasurface (the closed squares of Fig.\ \ref{fig:both}(b)) can also be understood from Fig.\ \ref{fig:both}(c), which reveals temporal enhancement of transmitted power. 

In contrast, the series-type metasurface drawn by the inset of Fig.\ \ref{fig:both}(d) can selectively absorb specific waveforms. Again, in Fig.\ \ref{fig:either} the individual capacitor-based metasurface temporarily stored the incoming energy in capacitors to dissipate it in resistors. However, this is now prevented by the electromotive force of the inductors. Likewise, long-pulse current cannot be absorbed by the inductor part, as the capacitors are fully charged up before the current reaches the inductor part. For these reasons, both a short pulse and CW are weakly absorbed as simulated in Fig.\ \ref{fig:both}(d). However, the series-type metasurface still absorbs some waveforms that were strongly absorbed by both of the individual metasurfaces. This is demonstrated by the closed squares of Fig.\ \ref{fig:both}(e), which combined the circuit configurations used in Figs.\ \ref{fig:either}(b) and \ref{fig:either}(e). Unlike the parallel-type metasurfaces, the absorptance value of the series-type metasurface at each pulse width is close to the \emph{smaller} value of the individual structures (cf.\ Figs.\ \ref{fig:either}(b) and \ref{fig:either}(e)). This waveform selectivity is experimentally realisable as plotted by the open squares of Fig.\ \ref{fig:both}(e), although circuit parasitics and superimposed direct current characteristics caused some difference. Similarly with the parallel case, this variation can be further increased by changing the time constants as seen from the circles of Fig.\ \ref{fig:both}(e) (now $C=$ 10 nF, $L=$ 10 $\mu$H, $R_c=$ 10 k$\Omega$, $R_l=$ 2 $\Omega$, $f_c=$ 57.3 MHz and $f_l=$ 45 MHz). The time domain response of such a series-type metasurface (the circles of Fig.\ \ref{fig:both}(e)) is plotted in Fig.\ \ref{fig:both}(f), which also supports temporal reduction of the transmitted power. 

\begin{figure}[t!]
\centering
\includegraphics[width=0.4\textwidth]{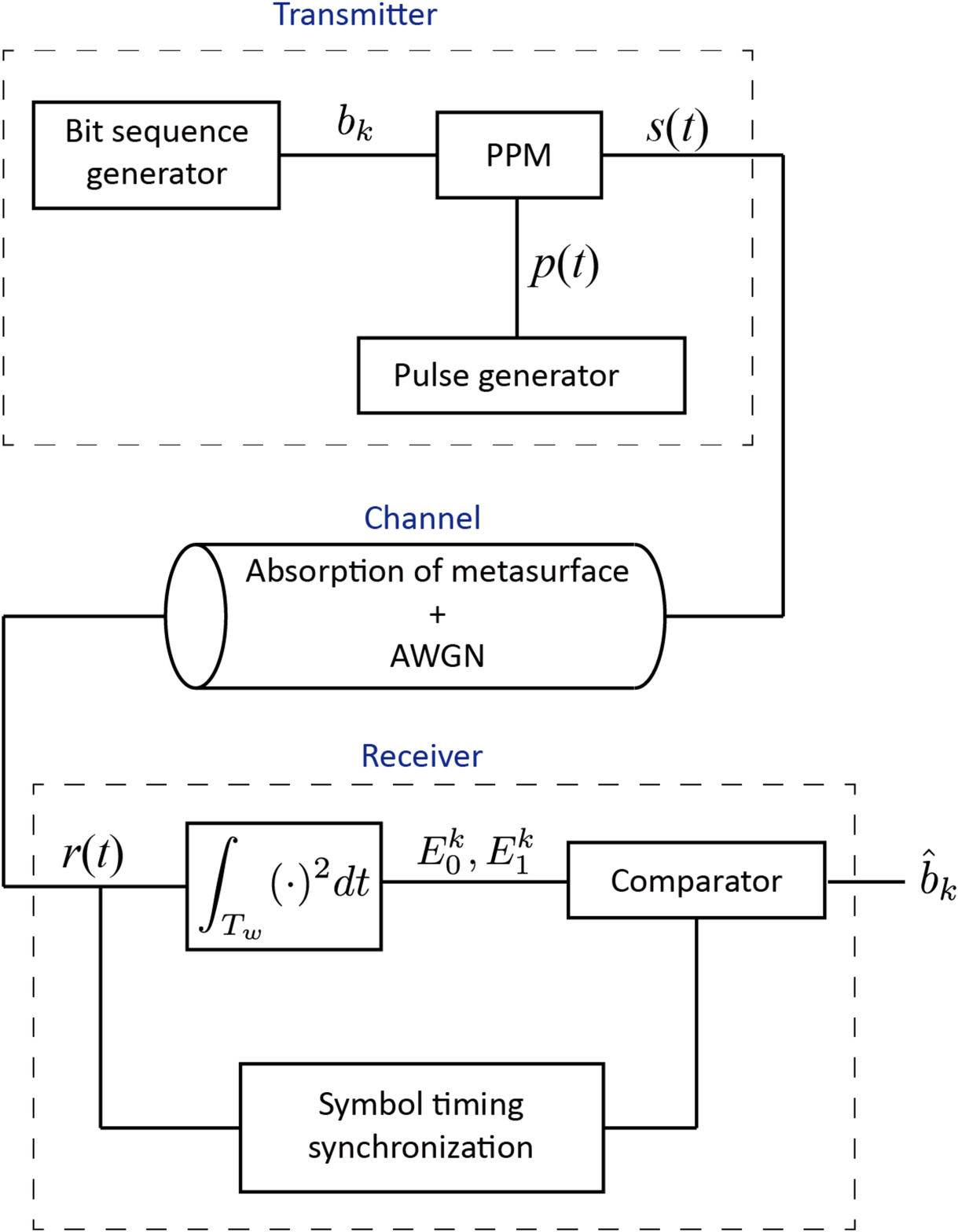}
\caption{Simulation overview of wireless communication performance evaluation.}
\label{fig:comm_system}
\end{figure}

These waveform selectivities are expected to develop new kinds of techniques and applications in electromagnetics, especially in wireless communications as demonstrated below. Here the structure of the transmitter is shown in Fig.~\ref{fig:comm_system}.
The transmitter uses a binary pulse position modulation (PPM) scheme,
in which bit information is sent as two pulse positions.
The block of the PPM controls a programmable time delay,
which determines when the pulse generator will be triggered.
Assuming $K$ as the total number of transmitted bits,
the PPM signal $s(t)$ can be expressed as
\begin{eqnarray}
s(t) = \sum_{k=1}^{K} \mbox{Re} \bigg[ p\bigg(t-\bigg(k-1\bigg)T_s -b_k T_s/2\bigg) e^{j(2\pi f_c t)} \bigg]
\label{eq:transmitted_signal}
\end{eqnarray}
\noindent
where $b_k$ is the $k$-th transmitted bit information,
namely, $b_k\in\{0,1\}$ ($k=1,2,\cdots,K$),
and $T_s$ is the symbol duration. Here, $p(t)$ represents the baseband signal, which is given by
\begin{equation}
p(t) = \begin{cases}
A & \mbox{if } 0 \leqq t \leqq T_w \\
0 & \mbox{otherwise,}
\end{cases}
\end{equation}
where $A$ and $T_w$ are the amplitude of the baseband signal and the pulse width, respectively.
Note that, in this paper, the bandwidth of the baseband signal $f_b = 1 / T_w$ is much smaller than the carrier frequency $f_c$ of 4.4 GHz (narrow-band condition).
Consequently, even if we set several kinds of the pulse width (we set $T_w$ to $20, 2000$ and $20000$ ns in this paper),
the transmitted signal can be assumed as if it has only a single frequency component of the carrier frequency.

Fig.~\ref{fig:comm_system} also shows the structure of the receiver.
Here, we pay attention to the energy detection (non-coherent detection) as the received detection scheme.
In the energy detection, since the binary PPM scheme chooses one from two location assignments in the $k$-th symbol,
we calculate two kinds of $k$-th energies for the corresponding pulse locations
from the received signal $r(t)$ at the two kinds of time durations as follows:
\begin{equation}
E_0^k = \int_{(k-1)T_s}^{(k-1)T_s+T_w} [r(t)]^2 dt
\label{eq:e_k0}
\end{equation}
\begin{equation}
E_1^k = \int_{(k-1)T_s+T_s/2}^{(k-1)T_s+T_s/2+T_w} [r(t)]^2 dt.
\end{equation}
\noindent
Comparing $E_0^k$ with $E_1^k$, the received bit information $\hat{b}_k$ can be decided as
\begin{eqnarray}
\hat{b}_k = \begin{cases}
0 & \mbox{if } E_0^k > E_1^k \\
1 & \mbox{otherwise.}
\end{cases}
\end{eqnarray}
\noindent
As seen from this equation, the PPM requires no threshold.
On the other hand, on-off keying (OOK) modulation, which is often used in digital communications with a pulse modulation scheme,
requires a threshold to distinguish signals and noise.
Because the threshold depends on the signal-to-noise power ratio (SNR), it is difficult to determine it in advance.
In this sense, the PPM is superior to the OOK modulation.
Furthermore, the symbol timing in Fig.~\ref{fig:comm_system} is synchronised with pilot signals.

\begin{figure}[t!]
\centering
\includegraphics[width=0.42\textwidth]{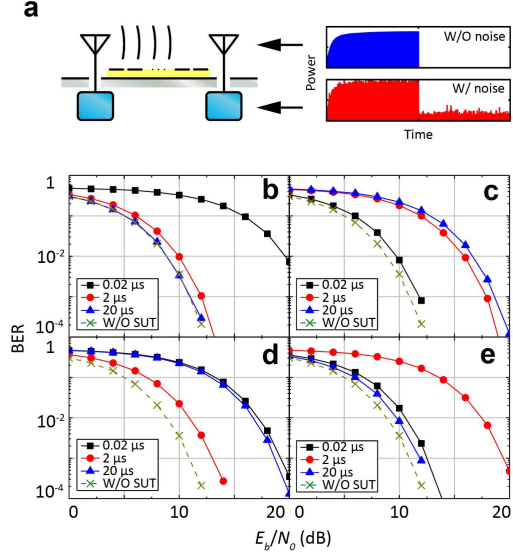}
\caption{\label{fig:comm} Waveform-selective wireless communications. (a) The overview of a wireless communication simulation where a surface wave propagates above metasurfaces. The signal received by an antenna (the blue line) (assumed as transmitted powers in TEM waveguide simulations) experiences the AWGN channel inside the demodulating circuit (the red line). (b)-(e) The BER performances for a simple narrow-band modulation, e.g.\ a PPM scheme, as a function of the energy per bit to the noise power spectral density ratio ($E_b/N_0$) in the communication system equipped with the capacitor-based, inductor-based, parallel-type and series-type metasurfaces, respectively. Optimal pulse widths for the best BER performances are determined by the waveform selectivity of the metasurfaces.}
\end{figure}

Fig.\ \ref{fig:comm} shows bit error rate (BER) performances using four-types of waveform-selective metasurfaces (Figs.\ \ref{fig:either}(b) and (e) and the closed circles of Figs.\ \ref{fig:both}(b) and (e), respectively, for Figs.\ \ref{fig:comm}(b), (c), (d) and (e)) together with transmitted powers calculated from TEM waveguide simulations as the received signals (i.e.\ as the blue line of Fig.\ \ref{fig:comm}(a)). These signals then experience the additive white Gaussian noise (AWGN) channel inside the demodulator (the red line). In Figs.\ \ref{fig:comm}(b) to (d) a signal becomes more erroneous due to the selective absorption of the metasurface, if the curve is shifted upwards from that without surface under test (SUT). For example, in Fig.\ \ref{fig:comm}(b), the 20-$\mu$s-long signal exhibited almost the same result as that without SUT, whereas the 0.02-$\mu$s-long signal showed reduced performance due to the waveform-selective metasurface. That is, in this case, the proposed waveform-selective communication can receive the signal with 20 $\mu$s pulse width and eliminate that with 0.02 $\mu$s pulse width. Note that even in these simulations our metasurfaces are not used as a band pass filter to eliminate low or high frequency components. From this perspective, it can be concluded that the waveform-selective wireless communication successfully receives wireless signals with arbitrary pulse widths even at the same carrier frequency.

Importantly, we employed none of special techniques, such as the code division multiple access (CDMA) and orthogonal frequency division multiple access (OFDMA) techniques \cite{hara1997overview, holma2009lte}, which basically make use of wideband spectrum characteristics. In other words, we just considered a simple narrow-band modulation technique. Nevertheless, the waveform-selective communication can distinguish the narrow-band signals even at the same carrier frequency of 4.4 GHz by adapting several kinds of pulse widths. Moreover, the waveform selectivity can be combined with other wireless communication techniques including the multiple access techniques such as CDMA and OFDMA. This idea can potentially solve a long-lasting problem in available radio frequency resources, which are limited by a growing demand on wireless communications \cite{schwarz2013pushing}, since the waveform selectivity allows us to effectively share even the same frequency resource by assigning different pulse widths. 

\begin{figure}[t!]
\centering
\includegraphics[width=0.25\textwidth]{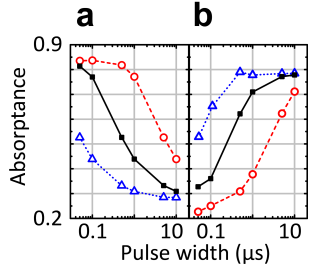}
\caption{\label{fig:timeConstants} Pulse width dependences of (a) a capacitor-based metasurface with various capacitances and (b) an inductor-based metasurfaces with various inductances. These metasurfaces used the same conditions as those of Figs.\ \ref{fig:either}(b) and \ref{fig:either}(e) except for the capacitance and inductance. The results using the default capacitance and inductance values are plotted with the solid curves. These values are respectively increased and decreased by a factor of 10 in the dashed curves and dotted curves. }
\end{figure}

The time constants are very important to determine how the waveform-dependent metasurfaces behave in response to the pulse width of an incoming wave. This is demonstrated in Figs.\ \ref{fig:timeConstants}(a) and \ref{fig:timeConstants}(b), where the time constants of the metasurfaces used in Figs.\ \ref{fig:either}(b) and \ref{fig:either}(e) were varied, respectively. The dashed lines increased the default capacitance or inductance (the solid) by a factor of 10, while the dotted lines decreased the default values by a factor of 10. For simplicity, in these simulations the only value changed was either the capacitance or inductance, although realistically these changes have influence over the self-resonant frequencies. As a result, it turned out that increasing the capacitance and inductance (i.e.\ increasing the time constants $R_cC$ and $L/R_l$) led to shifting the original curves (the solid) to the right, since these changes made the capacitors store more incoming energy and the inductors maintain the electromotive force longer. On the other hand, the reductions of the capacitance and inductance resulted in shifting the original curves to the left. This is because the capacitors store less energy and the inductors maintain the electromotive force shorter. It is very important to understand from these figures that these pulse width dependences are readily controllable by changing the time constants only, as long as the necessary time constant values (i.e.\ the actual circuit components) are available. 

In summary we have demonstrated a new concept of metasurfaces, which selectively absorb or transmit specific waveforms even at the same frequency. Similarly with other metasurfaces \cite{lier2011octave,sun2012gradient}, these waveform-selective metasurfaces can be deployed not only as coating of ordinary conducting surfaces but also as part of devices, which gives them another functionality to control electromagnetic waves. For example, integrating these metasurfaces with antennas leads to sensing specific signals, as our simulation demonstrated that they can vary BER performances in accordance with pulse widths.



\providecommand{\noopsort}[1]{}\providecommand{\singleletter}[1]{#1}%


\clearpage
\appendix
\textbf{{\Large Supplementary Information}}
\\\\Waveform Selectivity at the Same Frequency 
\\Hiroki Wakatsuchi, Daisuke Anzai, Jeremiah J. Rushton, Fei Gao, Sanghoon Kim, and Daniel F. Sievenpiper

\textbf{\begin{itemize}
\item Rectification
\item Modelling
\item Simulation method
\item Measurement method
\item Low power responses
\item Voltage in a capacitor and current in an inductor
\item Metasurface with only diodes and resistors
\item Theoretical performance of PPM Transmission
\end{itemize}
}

\begin{table*}[b!]
\caption{Fourier component of various modes for half and full wave rectifications}
\label{tab:rect}
\begin{center}
\begin{tabular}{c|p{3cm}p{3.8cm}p{3.8cm}}
Fourier term & W/O rectification: & Half wave rectification: & Full wave rectification:\\
 & $\cos$ & ($\cos$+$|\cos|$)/2 & $|\cos|$\\
\hline
0&0&1/$\pi$&2$\pi$\\
$f$ &1/2&1/4&0\\
2$f$ &0&1/(3$\pi$)&2/(3$\pi$)\\
3$f$ &0&0&0\\
4$f$ &0&-1/(15$\pi$)&-2/(15$\pi$)\\
\hline
\end{tabular}
\end{center}
\end{table*}
\section*{Rectification}
Microwave diodes rectify incoming signals to a static field in the following manner. Surface waves can be represented by a cosine function $\cos{(2\pi ft)}$, where $f$ and $t$ are respectively the frequency and time. Note that for simplicity other variables including the spatial position, phase delay, magnitude, etc, are all omitted here. When the input signal is fourier-transformed (i.e.\ $\int^{\infty} _{-\infty}\cos{(2\pi ft)}e^{-j2\pi t}dt$), the output spectrum contains only $f$. 

If the surface current is rectified by a diode, however, the rectified signal becomes ($\cos{(2\pi ft)}+|\cos{(2\pi ft)}|$)/2, which through a fourier transform results in an infinite set of frequencies with decreasing magnitudes. The largest term is at zero frequency, or a static field. 

This rectification to a static field is further enhanced, if a full wave rectification is introduced as the metasurfaces demonstrated in the paper. In this case the incoming signal is rectified to $|\cos{(2\pi ft)}|$. All of these are summarised in Table \ref{tab:rect}. 

\section*{Modelling}
\begin{figure}[b!]
\centering
\includegraphics[width=0.3\textwidth]{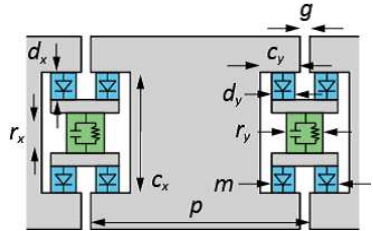}
\caption{\label{fig:detailedModel} The geometry simulated. The dimensions were given from $c_x=7.6$, $c_y=1.7$, $d_x=1.3$, $d_y=0.5$, $g=1.0$, $m=2.4$, $p=18.0$, $r_x=1.0$ and $r_y=2.0$ (all in mm). }
\end{figure}

Waveform-selective metasurfaces were modelled as described in Fig.\ \ref{fig:detailedModel}. Diodes were modelled by a SPICE model whose parameters are summarised in Table \ref{tab:spice}. The dielectric substrate used (Rogers3003) had relative permittivity of 3.0 and dielectric loss tangent of 0.0013. The relative permeability was 1.0. The conducting patches of the metasurfaces were modelled by 17-$\mu$m-thick copper, which had bulk conductivity of 5.8$\cdot$10$^7$ S/m, relative permittivity of 1.0 and relative permeability of 0.999991. 

\begin{table}[b!]
\caption{SPICE model parameters used for diode modelling}
\label{tab:spice}
\begin{center}
\begin{tabular}{p{2cm}p{1.5cm}p{1.5cm}}
Parameter & Units & Value\\
\hline
$B_V$ & V & 7.0\\
$C_{J0}$ & pF & 0.18\\
$E_{G}$ & eV & 0.69\\
$I_{BV}$ & A & 1$\cdot$10$^{-5}$\\
$I_{S}$ & A & 5$\cdot$10$^{-8}$\\
$N$ &  & 1.08\\
$R_{S}$ & $\Omega$ & 6.0\\
$P_{B}$ ($VJ$) & V & 0.65\\
$P_{T}$ ($XTI$) &  & 2\\
$M$ &  & 0.5\\
\hline
\end{tabular}
\end{center}
\end{table}

\section*{Simulation method}
All the simulations were performed by electromagnetic/circuit co-simulation to evaluate the absorptance $A$, reflectance $R$ and transmittance $T$ of metasurfaces. First, as drawn in Fig.\ \ref{fig:models}(e), we calculated the scattering parameters through an electromagnetic simulator Ansys HFSS 15.0. The metasurface simulated here had lumped ports which were later connected to circuit components, such as diodes, in circuit simulations performed by a circuit simulator Ansoft Designer 8.0. Effectively, this is equivalent to directly connecting them to the metasurface in the electromagnetic simulation. Hence, all the scattering parameters and absorptance were calculated by the circuit simulator. 

\begin{figure}[t!]
\centering
\includegraphics[width=0.45\textwidth]{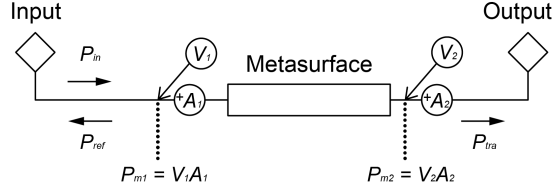}
\caption{\label{fig:circuitSim} Entire circuit configuration used for the circuit simulation. For the actual model we also connected lumped ports to the metasurface. }
\end{figure}

\begin{figure}[t!]
\centering
\includegraphics[width=0.25\textwidth]{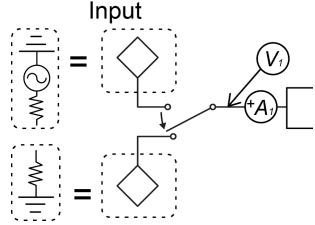}
\caption{\label{fig:switch} Introduction of switching system to produce pulses. In an arbitrary time the metasurface model was disconnected from the original input port and connected to a new port which did not have any excitation source. Each port had the same impedance as the port impedance of the metasurface model. }
\end{figure}

\begin{figure}[t!]
\centering
\includegraphics[width=0.45\textwidth]{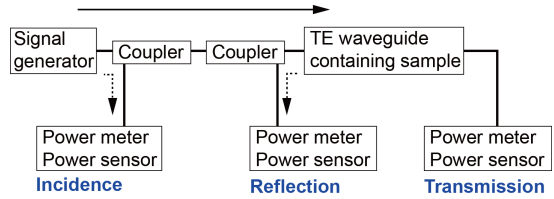}
\caption{\label{fig:measSys} Measurement system. }
\end{figure}

The absorbing performances in high power pulse and CW simulations were both evaluated in time domain using the electric circuit shown in Fig.\ \ref{fig:circuitSim}, where for simplicity we abbreviated lumped ports. In these simulations the input power was theoretically estimated by 
\begin{eqnarray}
P_{in}(t)=2P_{0}\sin^2{(2\pi ft)},\label{eq:PinCircuit}
\end{eqnarray}
\\where $P_{in}$ and $P_{0}$ are, respectively, the instant input power and the magnitude. Since the voltage and current meters next to the input port read power $P_{m_1}$ containing both the incident and reflected powers, the reflected power $P_{ref}$ was calculated by subtracting eq.\ (\ref{eq:PinCircuit}) from $P_{m_1}$ (i.e.\ $P_{ref}=P_{in}-P_{m1}$). For the transmitted power $P_{tra}$, the meters next to the output port were simply used (i.e.\ $P_{tra}=P_{m2}$), since there is no signal from the output port.  

For the pulse simulations, a switching system was inserted between the input port and neighboring meters as Fig.\ \ref{fig:switch}. This system played a role in switching the connection between the metasurface and input port. First, the metasurface model was connected to the original input port which generated a CW signal. Then, the metasurface model was disconnected from the original port and connected to a new port, which did not have any signal source. As a result, an arbitrary width of pulse was produced. The reflected and transmitted energies were then integrated over time and divided by the input energy to obtain the reflectance $R$ and transmittance $T$. These values were subtracted from one to calculate the absorptance $A$ (i.e.\ $A=1-R-T$). 

The high power CW simulations were more straightforward, i.e.\ the simulation was performed until a steady state, where the reflected and transmitted power were averaged and divided by the input power to obtain $R$ and $T$. There were a couple of factors to lead to instability in the simulations, for example, due to passivity and use of commercial diode models. For this reason some simulations were truncated earlier than others to avoid diverged results. 

\section*{Measurement method}
High power measurements were performed using the measurement system sketched in Fig.\ \ref{fig:measSys}. An input signal was generated from a signal generator (Agilent N5181A) and passed through waveguides, where metasurfaces were deployed. In addition, directional couplers (KRYTAR 102008010) were used to monitor the incident and reflected signals, which were measured by power sensors (Agilent N1921A). Another set of a power meter and power sensor was used after the waveguides to measure the transmitted power. In order to measure the scattering parameters as accurately as possible, the high power measurements were performed after several calibration steps. Some of the measurement devices were automatically controlled through LabVIEW software such that the input frequency could be quickly swept, which facilitated the measurement process substantially. A pulse width and duty cycle were changed with the signal generator and LabVIEW. 

Extra cares must have been paid on the pulse measurements. Power meters used (Agilent N1911A) had some operating modes, e.g.\ averaging mode and peak mode. Since our simulation results showed the same peaks for different pulse widths in time domain, we adopted the averaging mode. However, this indicates that depending on the duty cycle of the pulse, we needed to offset the received signals, e.g.\ if the received signal was 0 dBm and the duty cycle was 10 \% (i.e.\ -10 dB), then the actual magnitude was 10 dBm (i.e.\ 0 dBm - (-10 dB)). Due to the noise floor of the power meters, the duty cycle could not be set too small, otherwise the measured value would read the noise floor. This issue, however, needed to be compromised with another problems, e.g.\ the discharging time of stored energy in capacitors. Since the signal generator repeatedly produced pulses, the duty cycle was set long enough to ensure fully discharging the electric charges stored in capacitors. Because of these two issues we decided to set the duty cycle between 0.1 and 1 \% after some test measurements. 

Another point to note here is that the reflection was significantly small in the pulse measurements. Hence, the absorptance was estimated from the transmittance only, i.e.\ $A=1-T$. This fact can be numerically confirmed from Figs.\ \ref{fig:either}(c), \ref{fig:either}(f), \ref{fig:both}(c) and \ref{fig:both}(f) as well, where the reflected powers of all the waveform-selective metasurfaces were very limited, compared to the transmitted powers. 

\section*{Low power responses}
\begin{figure}[b!]
\centering
\includegraphics[width=0.45\textwidth]{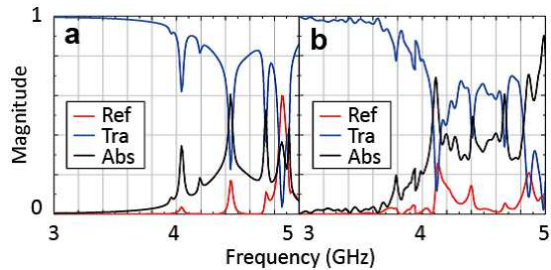}
\caption{\label{fig:lowPcap} Low power scattering profile of the capacitor-based metasurface in (a) simulation and (b) measurement.  }
\end{figure}

Low power scattering profiles of the capacitor-based metasurface are shown in Fig.\ \ref{fig:lowPcap} \cite{wakatsuchi2013waveform}. Since the input power was not large enough to turn on the diodes, the metasurfaces behaved as normal conducting surfaces and transmitted the most energy, especially up to 4.2 GHz in the simulation and up to 4.0 GHz in the measurement. The differences can be attributed to parasitics in the circuit elements, especially in the diodes, which were not fully modelled in the simulation for the sake of simplicity. 

\begin{figure}[t!]
\centering
\includegraphics[width=0.25\textwidth]{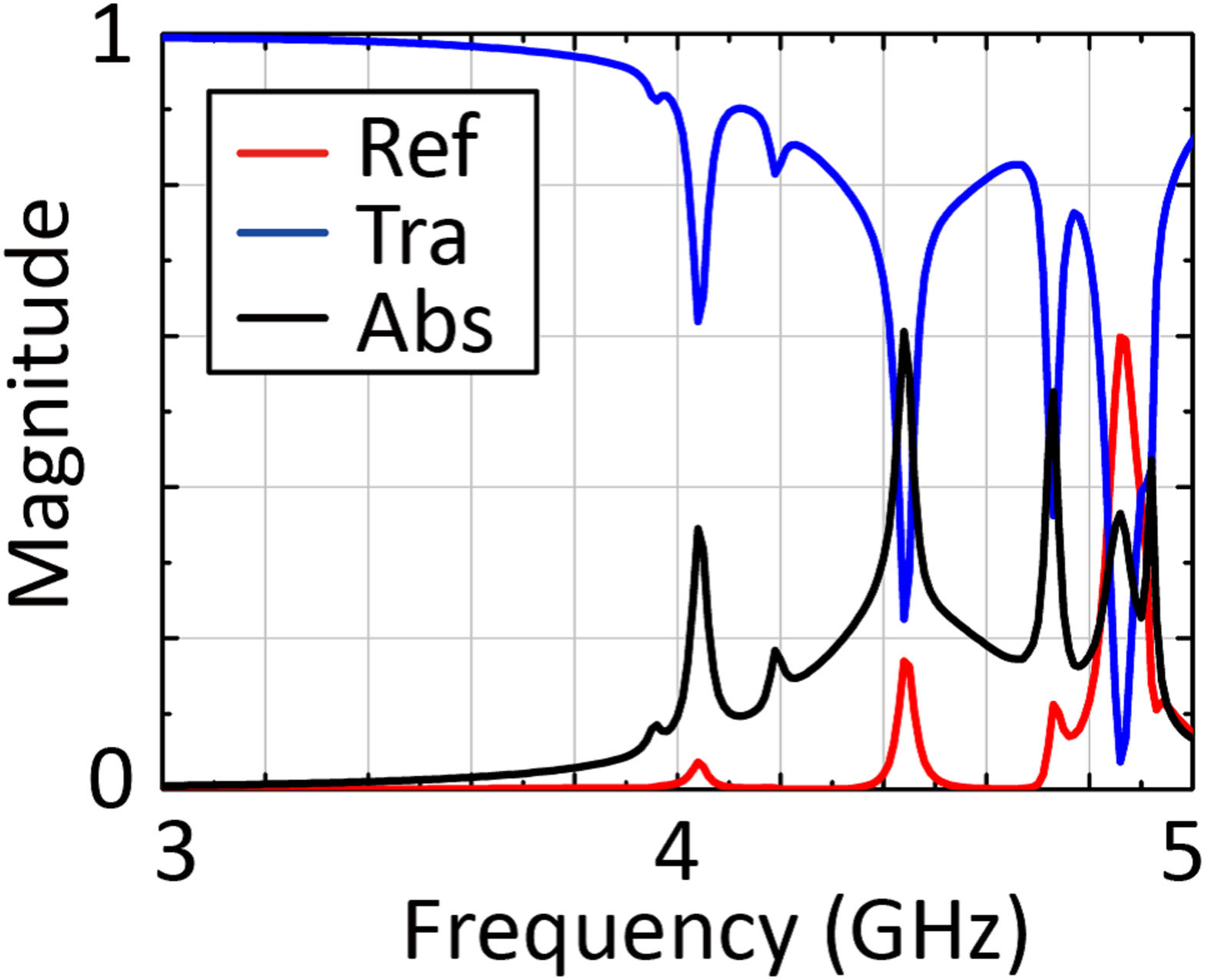}
\caption{\label{fig:lowPindScat} Low power scattering profile of the inductor-based metasurface in simulation.  }
\end{figure}

Since the low power responses do not depend on what components are used within diode bridges, the simulation result remains the same even if pairs of a parallel capacitor and resistor are replaced with pairs of a series inductor and resistor as an inductor-based metasurface. This is demonstrated in Fig.\ \ref{fig:lowPindScat}, which numerically shows the low power scattering profile of the inductor-based metasurface simulated in Fig.\ \ref{fig:either}(d). 

\section*{Voltage in a capacitor and current in an inductor}
\begin{figure}[t!]
\centering
\includegraphics[width=0.45\textwidth]{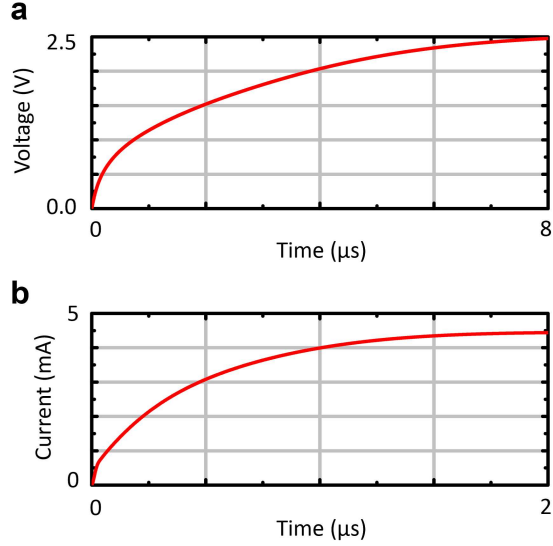}
\caption{\label{fig:vAndI} Voltage and current in metasurfaces. The capacitor-based metasurface simulated in Fig.\ \ref{fig:either}(b) gradually increases the voltages across the capacitors as in (a), which plots the voltage of the second capacitor from the front. On the other hand, the inductor-based metasurface simulated in Fig.\ \ref{fig:either}(e) permits more current to come in as (b), which represents the current flowing into the second inductor. }
\end{figure}

In Figs.\ \ref{fig:models}(a) and \ref{fig:models}(b) we explained how a capacitor and inductor respond to a rectified incoming wave in time domain. Such behaviours are numerically demonstrated in Figs.\ \ref{fig:vAndI}(a) and \ref{fig:vAndI}(b). These figures respectively plot the voltage across the second capacitor of the capacitor-based metasurface used in Fig.\ \ref{fig:either}(b) and the current flowing into the second inductor of the inductor-based metasurface used in Fig.\ \ref{fig:either}(e). As seen in these figures, the capacitor used in the capacitor-based metasurface is gradually charged up, while the inductor-based metasurface allows more current to come in, since the electromotive force is disappearing. 

\section*{Metasurface with only diodes and resistors}
\begin{figure}[t!]
\centering
\includegraphics[width=0.45\textwidth]{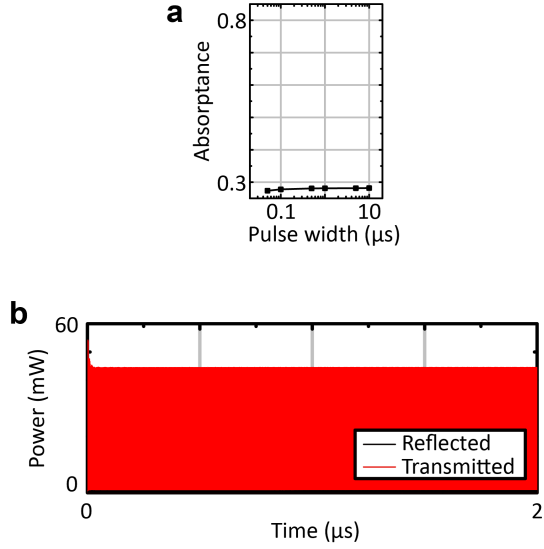}
\caption{\label{fig:withResiOnly} (a) Pulse width dependence and (b) time-domain response of the metasurface used in Fig.\ \ref{fig:either}(b) but without capacitors, namely with only diodes and resistors. }
\end{figure}

Capacitors and inductors play very important roles to create the waveform selectivity. This is confirmed from Fig.\ \ref{fig:withResiOnly}, which shows the pulse width dependence and time domain response of the metasurface used in Fig.\ \ref{fig:either}(b) but without the capacitors, namely with only diodes and resistors. Under this circumstance the metasurface no longer exhibits any waveform dependence, because there is no circuit element to temporarily control the incoming energy.  

\section*{Theoretical performance of PPM transmission}
\begin{figure}[t!]
\centering
\includegraphics[width=0.25\textwidth]{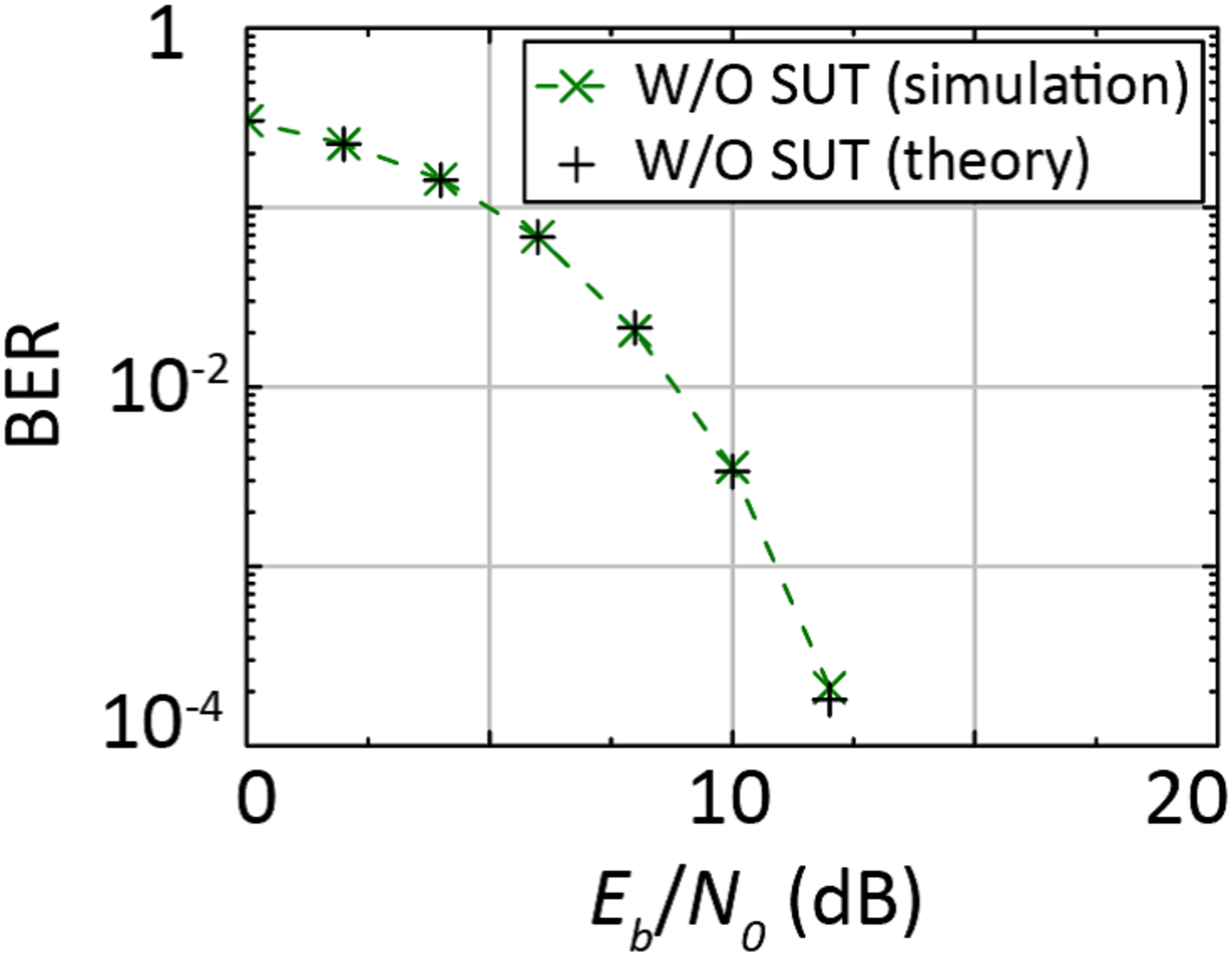}
\caption{\label{fig:theory} Theoretical analysis and simulation result for the PPM transmission with the energy detection under the AWGN channel. }
\end{figure}

This section explains the theoretical wireless communication performance, i.e.\ bit error rate (BER) performance, for the PPM transmission under the additive white Gaussian (AWGN) channel.
The average BER for the binary PPM transmission with the energy detection can be theoretically analysed as
\begin{equation}
P_b(E_b/N_0) = \frac{1}{2} \exp\left(-\frac{E_b}{2N_0} \right)
\label{eq:error_single}
\end{equation}
where $E_b/N_0$ indicates the energy per bit to the noise power spectrum density ratio.
Fig.~\ref{fig:theory} shows the theoretical BER performance analysed from the above equation.
Furthermore, this figure also includes the result evaluated from the computer simulation without any metasurface (i.e.\ just an empty TEM waveguide).
As seen in this figure, these two results agree with each other very well. 
This means that the computer simulation for the wireless communication performance evaluation has been properly performed, and the overall BER results in this paper have a certain level of reliability.

\end{document}